\RequirePackage{lineno}
\documentclass[aps,prx,twocolumn,superscriptaddress]{revtex4}
\usepackage{graphicx}
\usepackage{amsmath}
\usepackage{color}

\usepackage{mathtools}

\DeclarePairedDelimiterX\braket[2]{\langle}{\rangle}{#1 \delimsize\vert #2}

\graphicspath{{./}}
\begin{document}

\setpagewiselinenumbers
\modulolinenumbers[5]

\title{Evolution of topological edge modes from honeycomb photonic crystals to triangular-lattice photonic crystals}

\author{Jin-Kyu Yang}
\email[]{jinkyuyang@kongju.ac.kr}
\affiliation{Department of Optical Engineering, Kongju National University, Cheonan, 31080, South Korea.}
\affiliation{Institute of Application and Fusion for Light, Kongju National University, Cheonan, 31080, South Korea.}
\author{Yongsop Hwang} 
\affiliation{Institute of Application and Fusion for Light, Kongju National University, Cheonan, 31080, South Korea.}
\author{Sang Soon Oh}
\email[]{OhS2@cardiff.ac.uk}
\affiliation{School of Physics and Astronomy, Cardiff University, Cardiff CF24 3AA, United Kingdom.}

\date{\today}
\begin{abstract} 
The presence of topological edge modes at the interface of two perturbed honeycomb photonic crystals with $C_6$ symmetry is often attributed to the different signs of Berry curvature at the K and K$'$ valleys.
In contrast to the electronic counterpart, the Chern number defined in photonic valley Hall effect is not a quantized quantity but can be tuned to finite values including zero simply by changing geometrical perturbations. 
Here, we argue that the edge modes in photonic valley Hall effect can exist even when Berry curvature vanishes. 
We numerically demonstrate the presence of the zero-Berry-curvature edge modes in triangular lattice photonic crystal slab structures in which $C_3$ symmetry is maintained but inversion symmetry is broken. 
We investigate the evolution of the Berry curvature from the honeycomb-lattice photonic crystal slab to the triangular-lattice photonic crystal slab and show that the triangular-lattice photonic crystals still support edge modes in a very wide photonic bandgap. 
Additionally, we find that the edge modes with zero Berry curvature can propagate with extremely low bending loss.

\end{abstract}

\maketitle

\section{Introduction}
Topological insulators, which are insulating in the bulk part while conductive along the edge, have been intensively studied due to their intriguing physical properties as well as potential applications~\cite{hasan2010colloquium}.
As an optical counterpart of the topological insulators in condensed matter physics, photonic topological insulators (PTIs) have been proposed and demonstrated in optical systems~\cite{khanikaev12, rechtsman12}.
PTIs show unique characteristics, for example, the guided modes along the edge or interface of PTIs which are robust against defects and deformations due to topological protection~\cite{chen14, cheng16, ma2016all, chen17}.
Such robustness has been demonstrated at telecommunication wavelengths~\cite{shalaev2019robust}.

Recently, optical quantum spin-Hall effect (QSHE) and optical quantum valley-Hall effect (QVHE) have been realized by introducing geometrical perturbation in  PTIs such as a honeycomb (HC) photonic crystal (PhC) structure~\cite{wu2015honeycomb, chen17, ma2016all}. 
For  the QSHE PTIs~\cite{wu2015honeycomb}, a photonic bandgap (PBG) is created at Dirac point by making perturbations (extend, shrink) in a way that the $C_6$ symmetry is maintained. 
As a result, the pseudo-time-reversal symmetry is protected and pseudo-spin channels are maintained (pseudo-time-reversal operator $\mathcal{T}$ is defined to satisfy $\mathcal{T}^2 = -1$). 
For the QVHE PTIs, the inversion symmetry breaking, as in a staggered HC structure~\cite{he2019silicon, Meh2020optica} and  kagome lattice ~\cite{wong2020kagome}, opens a PBG. For both QSHE and QVHE PTIs, edge modes, which are localized at the interface between two PTIs with different signs of perturbation, exist in the PBGs.

Although the near 100\% transmission have been shown using the edge modes in staggered HC~\cite{wu2015honeycomb} and kagome lattices~\cite{wong2020kagome}, the origin of the edge modes in optical QVHE is not clearly understood because they are not topologically protected as in the Chern insulators (Chern number is zero for optical QVHE). 
One way of explaining the existence of edge modes in optical QVHE is the bulk-edge correspondence in an extended parameter space ($k$, $a_g)$ where $k$ is the wavevector and $a_g$ is the perturbation strength~\cite{saba2020perturb}. 
Another most common explanation is the valley degree of freedom which originates from the different signs of Berry curvature for different valleys. 
However, it is not clear whether the different signs of Berry curvature for different valleys is a necessary condition for the existence of the edge modes and reflection-less propagation at the bending. 

In this paper, we report that edge modes can be created even when the Berry curvature vanishes. 
We demonstrate this by studying the Berry curvature of photonic bands for PhC slabs with the staggered HC lattices including the HC lattice and the triangular lattice. 
We show the triangular-PhC (Tri-PhC) is an extreme case of the honeycomb PhCs (HC-PhCs) which holds their topological characteristics because the reduction of one hole in the unit cell to null leads to the Tri-PhCs. 
This argument is supported by the calculation of photonic band structures and Chern numbers and the simulation of one-way propagation.

\section{Photonic band structure analysis}

\begin{figure}[t]   
\includegraphics[width=0.5\textwidth]{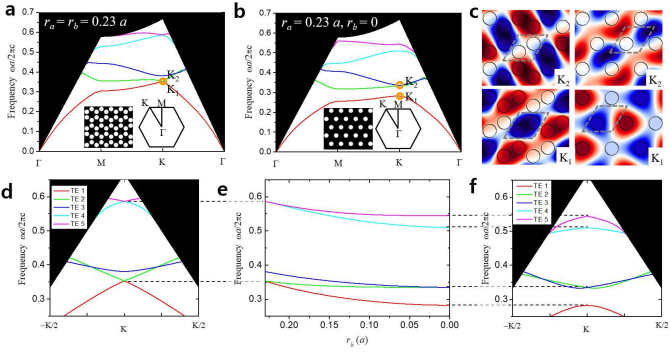}
\caption{ 
    Characteristics of the TE photonic band structures.
    (a) Band diagram of the air-hole HC-PhC slab structure with  $r_a=r_b=0.23a$ 
    (b) Band diagram of the air-hole Tri-PhC slab structure with $r_a=0.23a$ ($r_b=0$) 
    (c) Magnetic field profiles of the normal component ($H_z$) at the center of the slab. K$_1$, K$_2$ means the first and the second K-point TE modes shown in (a, b). 
    (d) Band diagram near the K point at  HC-PhC slab
    (e) Frequency of the K-point TE modes as a function of $r_b$.
    (f) Band diagram near the K point at the Tri-PhC slab
  }
  \label{band}
\end{figure}

First, we perform photonic band structure analysis to understand the characteristics of the guided modes in the staggered HC-PhC slab structure composed of two air holes in its unit cell. 
It is worth noting that the structure becomes a HC-PhC slab (the inset of Fig.~\ref{band}(a)) if the radii of the two holes are identical and it becomes a Tri-PhC slab (the inset of Fig.~\ref{band}(b)) if one of the air holes is missing. 
Therefore, the geometrical transition from the HC-PhC to the Tri-PhC can be described by the change of the radius of the smaller air hole ($r_b$) from $0.23a$ to zero.

Figure~\ref{band}(a) shows the band diagram of the HC-PhC slab ($r_b=0.23a$) calculated by the three-dimensional (3-D) plane wave expansion method~\cite{MPB}.
Here, the refractive index of the slab is set as 3.16, and the radius of the large air hole ($r_a$) and the thickness of the slab ($t$) are fixed as $0.23a$ and $0.47a$, respectively.
At the K point in the band diagram of the HC-PhC, one can observe the Dirac cone, where the lowest and the second-lowest TE bands meet with a linear slope.
In general, the Dirac cone exists when there are $C_3$ symmetry and the inversion symmetry with respect to the mid-point of two air holes in the unit cell (HC lattice)~\cite{wu2015honeycomb}. 

Figure~\ref{band} (b) shows the band diagram of the Tri-PhC structure ($r_b=0$).
When the inversion symmetry is broken (staggered HC lattice), the degeneracy at the Dirac cone is lifted opening a PBG where a one-way propagation mode can be introduced at the interface of two PTIs with different topological invariants such as valley Hall Chern numbers~\cite{ma2016all}.
The evolution of the band diagram by reducing the radius of one air hole is shown in Figs.~\ref{band}(d)-(f).
As the radius of the one air hole decreases, the PBG at K point opens because of inversion symmetry breaking and becomes wider until the smaller air hole is completely removed.

The PBG opening also can be numerically understood by the electromagnetic field profiles of the lowest and the second-lowest TE bands .
When the inversion symmetry is maintained, i.e. $r_a=r_b$, the magnetic fields are localized equally at both holes in one unit cell  (Fig. ~\ref{band}(c)) resulting in a degeneracy as shown in Fig. ~\ref{band}(a). 
However, the inversion symmetry breaking makes the magnetic field distribution asymmetric with respect to the midpoint of two holes for the first and band edge modes (K$_1$ and K$_2$). 
This is clear because the magnetic field is localized around larger holes  for K$_1$ in the first band of staggered HC-PhC slabs while the magnetic field is  localized around smaller holes for K$_2$ in the second band as shown Fig.~\ref{fieldV} (see Supplementary Information for more details). 
Remarkably, the overall field distributions of the two band edge modes are maintained until $r_b=0$ (Tri-PhC) which has the maximum PBG at the K point as shown in Fig.~\ref{band}(c). 
It is worth to mention the PBG opens even with a very small difference between $r_a$ and $r_b$ because the PBG originates from the structural inversion symmetry breaking.

\section{Berry curvature and valley Chern number}

\begin{figure}[tbp!]
  \includegraphics[width=0.48\textwidth]{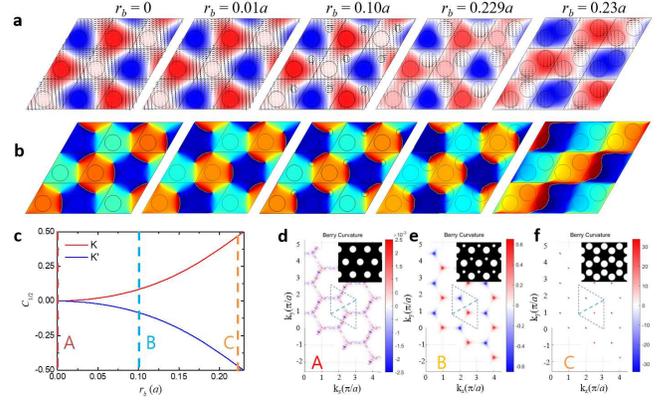}
  \caption{(a) The normal magnetic field ($H_z$) profiles of the first TE mode at the K point with the in-plane electric vectors ($\mathbf{E}_{\parallel}$) as the structure is gradually changed from the Tri-PhC ($r_b=0$) to the HC-PhC ($r_b=0.23a$) and (b) the phase of the in-plane electric vectors. (c) The half Chern number integrated in the half of the first Brillouin zone as marked in (d) $\sim$ (e) with dashed lines. (d-e) Berry curvatures of the first TE band (d) $r_b=0$ (A) (e) $r_b=0.10a$ (B), and (f) $r_b=0.229a$ (C, slightly perturbed HC-PhCs). }
  \label{chern}
\end{figure}

In order to verify the optical QVHE in the HC-PhCs with non-identical air holes, we investigate the evolution of the Berry curvature of the first band and its half Chern number from $r_b=0$ to $r_b=r_a$.
The Berry curvature $\mathcal{F}(\mathbf{k})=\nabla_\mathbf{k} \times i \braket{ \mathbf{E}_\mathbf{k} }{ \nabla_\mathbf{k} \mathbf{E}_\mathbf{k} }$  is numerically calculated by summing up the phases of the electric fields $\mathbf{E}_\mathbf{k}$ at the four points of the plaquette in the discretized $\mathbf{k}$ space. 
Then the valley Chern number $C_v$ is given as $C_v = C_{1/2,\mathrm{K}} - C_{1/2,\mathrm{K'}}$  where $C_{1/2,\mathrm{K}}$ ($C_{1/2,\mathrm{K'}}$) is the half Chern number calculated by integrating the Berry curvature over the triangular area around point K (K$'$) points~\cite{wong2020kagome}. 
Here, we consider two-dimensional (2-D) cylindrical air-hole PhCs because the field profile of the TE mode calculated by 2-D calculations is the same as the 3-D calculation except the variation along the plane-normal direction~\cite{comsol}. 
In consideration of the finite thickness of the PhC slab, the refractive index of dielectric materials in 2-D calculation is set as 2.4.  

Figures ~\ref{chern} (a, b) show the normal magnetic field component ($H_z$) profiles of the first TE mode at the K point, and its phase change. 
In Fig.~\ref{chern}(a), the arrows indicate the in-plane electric vectors.
At $r_b=0$ (Tri-PhC), the magnetic field is localized near $r_a$, and this localization behavior is preserved unless $r_b=r_a$. 
Another chiral property induced by breaking the inversion symmetry are found in the electric vectors around one of $r_a$. 

When the inversion symmetry is broken, there are two vortices one at the centre of hexagons and the other at the centre of smaller holes. In the amplitude plots, the amplitude becomes zero at the two vortices. In the phase plots, two vortices show different signs. This relation has been proved mathematically~\cite{Liu2016vortices}.

Interestingly, at $r_b=0$ (Tri-PhC), the Berry curvature is zero, however, two vortices with different signs still remain, which implies that the topologically protected mode could exist in the interface or edge within a wide spectral range of PBG of the Tri-PhCs (See Figs. ~\ref{shift1} and   ~\ref{shift2} in Supplementary Information).  
Recently, it was reported that inherently, photonic 2D Su-Schrieffer-Heeger (SSH) lattice has $C_{4v}$ point group symmetry, and zero Berry curvature~\cite{Liu2017zeroBerry, Liu2018zeroBerry}. 
According to Ref. ~\cite{Liu2017zeroBerry} the non-trival topological properties with zero Berry curvature could be realized by the curl of the magnetic field. 
The Tri-PhC has also $C_{6v}$ point group symmetry and zero Berry curvature that could support the edge state with one-way propagation because of the chiral property of the magnetic field.      

\section{One-way propagation}

\begin{figure}[tbp!]
  \includegraphics[width=0.48\textwidth]{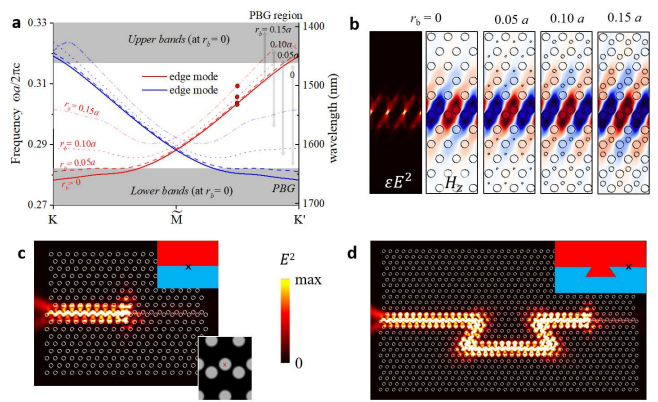}
  \caption{
    Analysis of photonic topological edge modes
    (a) Dispersion properties of the edge modes at the PTIs with various $r_b$. The gray region indicates the bulk bands. The white region (or the gray arrows at the right) indicates the PBG region of the given $r_b$. 
    (b) $H_z$ field profiles of the edge modes at the PTIs with various $r_b$ as marked with the red circles at (a). 
    (c) Time-averaged electric field intensity distribution of the edge mode at $\lambda = 1550 nm$ excited at the center of air holes in the straight interface as shown in the inset with the red mark. 
    (d) Time-averaged electric field intensity distribution in the {\rotatebox[origin=c]{180}{$\Omega$}} -shape interface. 
   The marks, ‘$\times$’ in the insets of (c) and (d) indicate the position of the excited clock-wise chiral source.}
  \label{interface}
\end{figure}

Given the continuity of the evolution of the band diagram, it is obvious that the Tri-PhC is one extreme case of the PTIs based on the staggered HC-PhCs.
Accordingly, the Tri-PhCs are expected to have topologically protected edge modes which are the same kind as the ones in the staggered HC-PhCs.
Hence, we designed a structure of a pair of the HC-PhCs with non-identical air holes, and one of them is vertically flipped and laterally shifted to form an interface between them as shown in Fig.~\ref{interface}.
To find guided modes along the interface, we calculated a band diagram in Fig.~\ref{interface}(a) showing the existence of edge modes.
The two guided modes with various $r_b$'s are clearly found in the PBGs which are denoted by the gray arrows at the right side of Fig.~\ref{interface}(a).
Figure ~\ref{interface}(b) shows the amplitude distributions of the magnetic field of the edge modes with different $r_b$'s marked with red circles in ~\ref{interface}(a). Because of the wide PBG, the edge mode in the Tri-PhCs ($r_b=0$) is strongly localized at the interface. 
However, as $r_b$ increases, the PBG becomes narrower and the localization of the edge mode in the interface becomes weaker. 

We also investigated the one-way propagation properties of the edge mode in the Tri-PhC by the 3-D Finite-Difference Time-Domain (FDTD) method\cite{Taflove}. In case of $r_b=0$ (Tri-PhC slab), it is clear that the edge mode propagates unidirectionally along the straight interface as shown in Fig.~\ref{interface}(c). Here, the mode is excited by the clockwise (CW) chiral source generated by two dipole sources with $\pi/2$ phase difference at $\lambda = 1550 nm$. In Fig.~\ref{interface}(c), the upper inset shows a schematics of a pair of Tri-PhC slab, and the lower shows the position of the chiral source with the red ‘$\times$’ mark. Even in the {\rotatebox[origin=c]{180}{$\Omega$}}-shape interface with four 120$^\circ$ bending geometry, the edge mode propagates along the interface without reflection near the sharp corners shown in Fig.~\ref{interface}(d).          

\begin{figure}[tbp!]
  \includegraphics[width=0.48\textwidth]{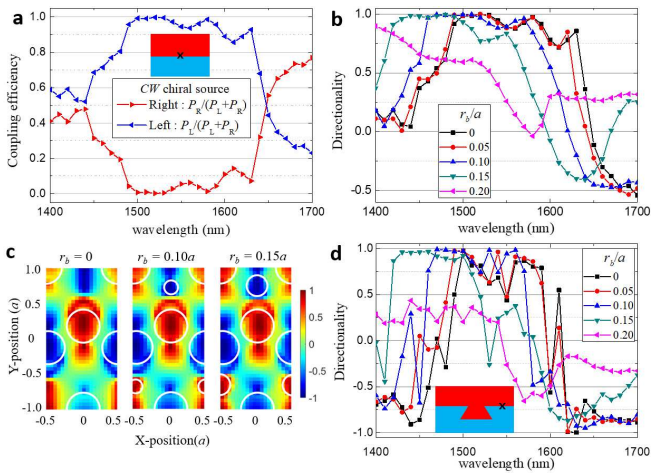}
  \caption{
    Characteristics of the one-way propagating edge modes. (a) One-way coupling efficiency of the CW chiral source into the edge modes in the straight interface with $r_b=0$ (Tri-PhC slabs).  (b) Directionality of the excited edge modes in the straight interface with various $r_b$. (c) The source-position dependent directionality with various $r_b$ at the CW chiral source. (d) Directionality of the excited edge modes in the {\rotatebox[origin=c]{180}{$\Omega$}}-shape interface with various $r_b$. 
}
  \label{map}
\end{figure}

For the quantitative analysis of chiral coupling of the guided mode along the PTI interface, we calculated one-way coupling efficiency defined as the ratio of the left (or right) propagating flux to both the left and the right propagating energy fluxes. 
Here, the CW chiral source was excited at the center of air holes, and the time-averaged Poynting flux monitored at the left (or right) edge of the photonic crystals was obtained for energy flux. 
Figure ~\ref{map}(a) shows one-way coupling efficiency as a function of wavelength of the CW chiral source excited at the center of the straight interface. The blue indicates the coupling efficiency of the CW chiral source into the propagating mode toward the left direction.
Because of the wide PBG of the Tri-PhCs, high one-way coupling efficiency over 90$\%$ is found in a broad spectral range from 1490 nm to 1600nm as shown in Fig. ~\ref{fieldstr}.
For the systematical analysis for the unidirectional coupling efficiency of the edge propagation, we defined the directionality as ($P_\textrm{L}-P_\textrm{R})/(P_\textrm{L}+P_\textrm{R})$ at the CW chiral source~\cite{he2019silicon}. 
Figure ~\ref{map}(b) shows the directionality in the HC-PhCs with non-identical air holes from $r_b/a=0$ to $r_b/a=0.20$ as a function of the wavelength. In case of $r_b=0$ (Tri-PhC slabs), the spectral range of directionality over 0.90 is about 60 nm. This wide spectral response of the directionality is very robust to the $r_b$. Especially, at $r_b/a=0.1$, the flat top directionality over 0.98 is about 50 nm. The blueshift of the frequency range of high directionality is owing to the blueshift of the PBG range as shown in Fig.~\ref{band}(e). 
The pictures in Fig.~\ref{map}(c) indicate the source-position dependent directionality with various $r_b$ when the CW chiral source was excited near the interface. 
In the simulation, we fixed the wavelength of the excited source as 1500 nm, 1470 nm, and 1440 nm for $r_b/a$ = 0, 0.1, and 0.15, respectively, for the the highest directionality at Fig.~\ref{map}(b). 
Here, the white lines indicate the air-hole boundary. 
When the point-like chiral source lies inside the upper air hole ($r_a$), the unidirectional propagation occurs along the left, however, the opposite directional propagation also observed when the source lies inside the lower air hole ($r'_a$). 
The directional propagation occurs dominantly when the source is located at the air hole, which is found at different $r_b$. It implies that the unidirectionality is related with both the geometry of the large air hole ($r_a$) and the chirality of the source. 
The asymmetric structural unidirectionality was also reported in the glide PhC waveguide structure~\cite{Lodahl15gliderWG}.

To demonstrate the robustness of unidirectional topological transport along the interface without reflection at a sharp bending, we investigated directionality in the {\rotatebox[origin=c]{180}{$\Omega$}}-shape interface with various $r_b$. 
Figure ~\ref{map}(d) shows the directionality as a function of wavelength of the CW chiral source at the right side of {\rotatebox[origin=c]{180}{$\Omega$}}-shape interface. 
The spectral range of directionality over 0.8 is observed over 60 nm which is reduced in comparison with the straight interface. 
However, at the optimal wavelength the directionality in the {\rotatebox[origin=c]{180}{$\Omega$}}-shape interface is almost same as that in the straight interface. 
For example, in case of $r_b=0.10a$, the directionality at $\lambda =1470$ nm in the straight interface and the {\rotatebox[origin=c]{180}{$\Omega$}}-shape interface is 0.995, 0.989, respectively. 
Especially, This indicates that the topological transport is robust with reflection at the sharp bending corner, at least for the optimal condition. The sudden decrease of the directionality at the longer wavelength is due to the bandedge behaviors of the edge mode (see Supplementary Information).  

\section{Conclusion}
We have numerically demonstrated the evolution of the topological behavior from the perfect HC slab to the Tri-PhC slab. 
When one of the two air holes in the unit cell of the HC lattice gradually decreases, the Dirac cone at the K point disappears and the photonic bandgap opens because of inversion symmetry breaking. 
From the systematic investigation of the evolution of the Berry curvature from the HC-PhC to the Tri-PhC, we have shown that the topological behaviors are maintained even with zero Berry curvature at the Tri-PhC. 
We have numerically demonstrated the lossless one-way propagation of the edge mode in the straight and bending interfaces with various size of the smaller air hole. We believe that our analysis of the staggered HC-PhC slab will be useful for design of highly efficient platform of lossless photonic integrated circuits. 

\begin{acknowledgments}

This research was supported by the Basic Science Research Program through the National Research Foundation of Korea (NRF) funded by the Ministry of Science and ICT (2017R1A2B4012181, 2020R1A2C1014498). 
This work is part-funded by the European Regional Development Fund through the Welsh Government (80762-CU145 (East)).

\end{acknowledgments}


\widetext
\clearpage
\textbf{\large Supplementary Information: Evolution of topological photonic effect from triangular-lattice photonic crystals to honeycomb photonic crystals}
\makeatletter

\begin{center}
\textbf{Jin-Kyu Yang$^{1,2,*}$, Yongsop Hwang$^{2}$, Sang Soon Oh$^{3,\dag}$}

\textit{$^{1}$Department of Optical Engineering, Kongju National University, Cheonan, 31080, South Korea.}

\textit{$^{2}$Institute of Application and Fusion for Light, Kongju National University, Cheonan, 31080, South Korea.}

\textit{$^{3}$School of Physics and Astronomy, Cardiff University, Cardiff CF24 3AA, United Kingdom.}
\end{center}

\date{\today}

\setcounter{equation}{0}
\setcounter{figure}{0}
\setcounter{table}{0}
\setcounter{page}{1}
\makeatletter
\renewcommand{\theequation}{S\arabic{equation}}
\renewcommand{\thefigure}{S\arabic{figure}}
\renewcommand{\bibnumfmt}[1]{[S#1]}
\renewcommand{\citenumfont}[1]{S#1}
\setcounter{section}{0}

\section{Band analysis}
\begin{figure}[hbp!]
  \includegraphics[width=0.48\textwidth]{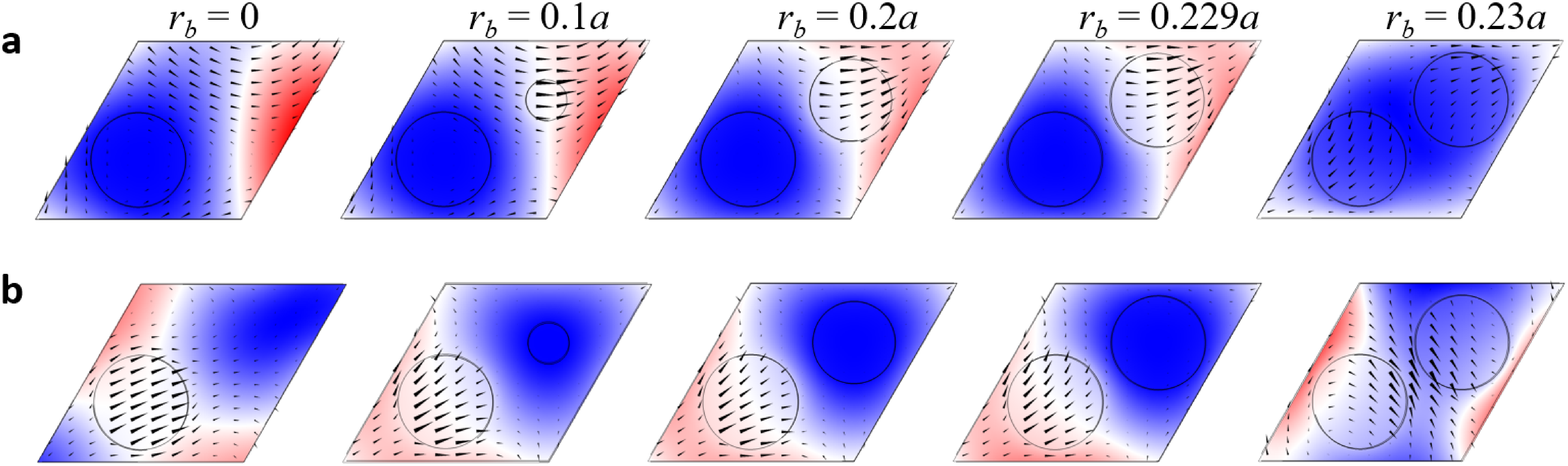}
  \caption{
The normal magnetic field ($H_z$) profiles with the in-plane electric vectors ($\vec{E}_{\parallel}$) as the structure is gradually changed from the Tri-PhC ($r_b=0$) to the HC-PhC ($r_b=0.23a$) (a)  The first TE mode at the K point.  (b) The second TE mode at the K point.
}
  \label{fieldV}
\end{figure}

  We investigated the field evolution to understand the origin of the photonic topological insulator (PTI) behaviors. Figure ~\ref{fieldV} shows the normal magnetic field profiles with the in-plane electric field vectors (arrows). The field profile of the first TE mode looks different from that of the second.  However, those modes are degenerate, which is caused by different propagation directionality, as shown in Fig.~\ref{band}(c) (areas in the gray dashed lines). 
Even in a very small structural perturbation regime, for example when $r_b=0.229a$, the magnetic profile changes dramatically because the structural symmetry is broken. This implies that the origin of the PTI behavior at the staggered HC-PhC slab structure is the symmetry breaking of the structure. Interestingly, the field profiles varies continuously without any jump in the field values until the extreme case, the Tri-PhC structure ($r_b=0$). This evolution behavior is also confirmed by the phase of the in-plane electric vectors as shown in Fig.~\ref{chern}(b).    

\begin{figure}[hbp!]
  \includegraphics[width=0.8\textwidth]{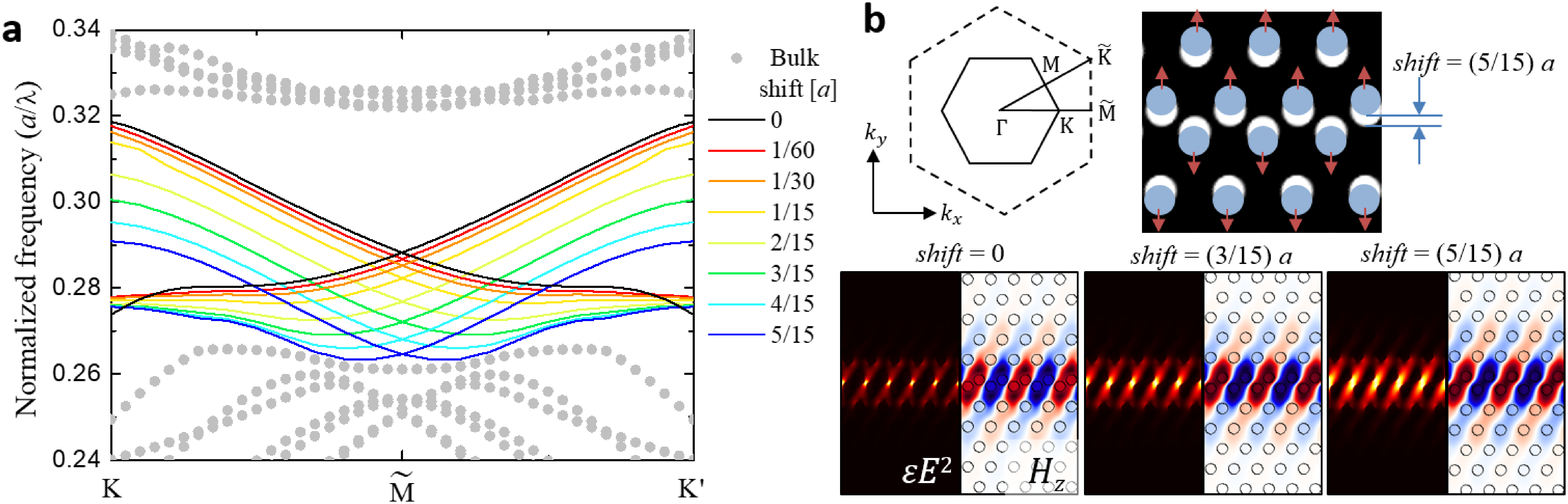}
  \caption{
Analysis of photonic topological edge modes at the interfacial gap along the normal direction of the interface in the Tri-PhC based topological structure. (a) Dispersion properties of the edge modes with various gap distances. (b) Schematic view of the creation of the interfacial gap. The left inset shows the Brillouin Zone and the symmetric point in reciprocal space.  (c) electric energy distributions($\epsilon E^2$) and the $H_z$ field profiles at the PTIs with various gap distances.    
}
  \label{shift1}
\end{figure}

 We numerically investigated the effect of the interfacial gap in the HC structure with $r_b=0$ (Tri-PhC). First, the interfacial gap is generated by shifting the patterns along the normal direction of the interface, oppositely.  Figure ~\ref{shift1}(a) shows the photonic band structures of the topological edge modes with various gap distances, and Fig.~\ref{shift1}(b) shows the schematic view of the shifted structure. As the gap increases, the normalized frequency of the photonic edge mode decreases due to the increase of the dielectric area near the interface. It is worth noting that if the gap increases further, the structure becomes close to the PhC glider waveguide structure. This means that PhC glide waveguide can be considered as a PTI structure originated from the symmetry breaking of the HC structure ~\cite{gliderWG} . 

\begin{figure}[tbp!]
  \includegraphics[width=0.8\textwidth]{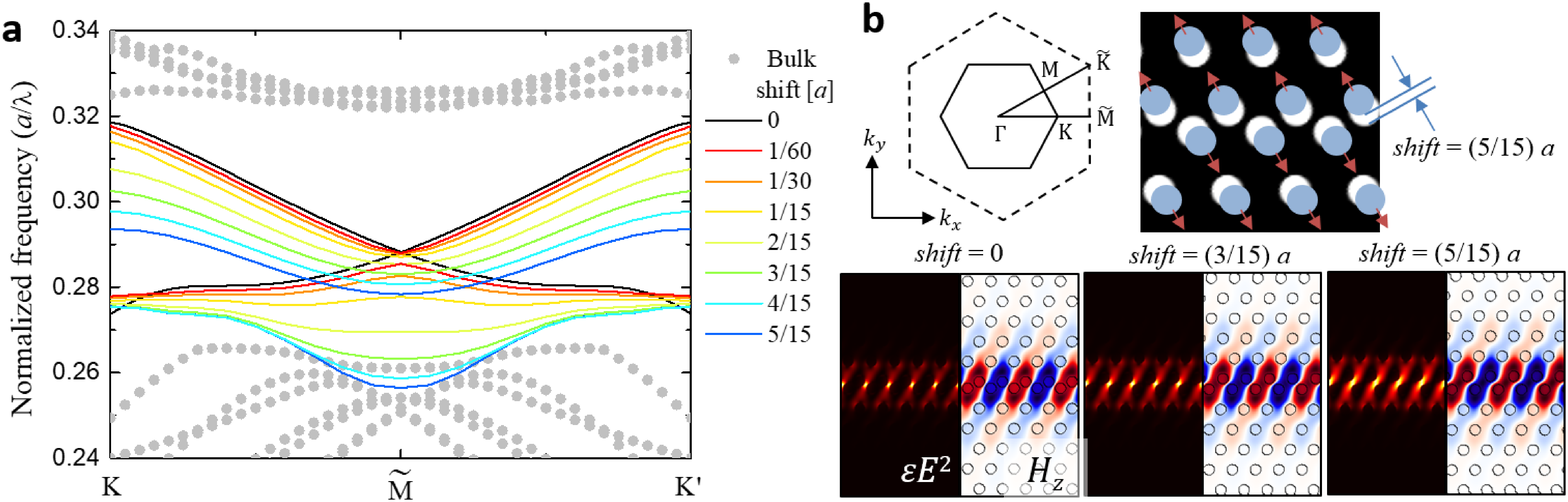}
  \caption{
Analysis of photonic topological edge modes with the diagonally-shifted interfacial gap in the Tri-PhC. (a) Dispersion properties of the edge modes with various gap distances. (b) Schematic view of the creation of the interfacial gap. The left inset shows the Brillouin Zone and the symmetric point in reciprocal space.  (c) electric energy distributions($\epsilon E^2$) and the $H_z$ field profiles at the PTIs with various gap distances.    
}
  \label{shift2}
\end{figure}

Second, we also investigated the dispersion of the photonic topological edge mode at the interfacial gap created by shifting the patterns along the 30$^\circ$.-direction of the interface, oppositely.  Figure ~\ref{shift2}(a) shows the dispersion curves of the photonic topological edge modes with various gap distances, and Fig.~\ref{shift2}(b) shows the schematic view of the 30$^\circ$.-directionally shifted structure. As the gap is formed asymmetrically, the mode splitting occurs due to the symmetry breaking. This implies that topological edge mode still have structural symmetry along the normal to the interface, which induce the Dirac-point-like band crossing of the two edge modes. However, if the interfacial symmetry is broken, no more Dirac cone exists, but the PBG opens. As the gap increases, PBG becomes wide, however, with further increase of the gap, PBG becomes narrow because of the decrease of the frequency of the edge mode.  

\section{Field analysis}

\begin{figure}[hbp!]
  \includegraphics[width=0.8\textwidth]{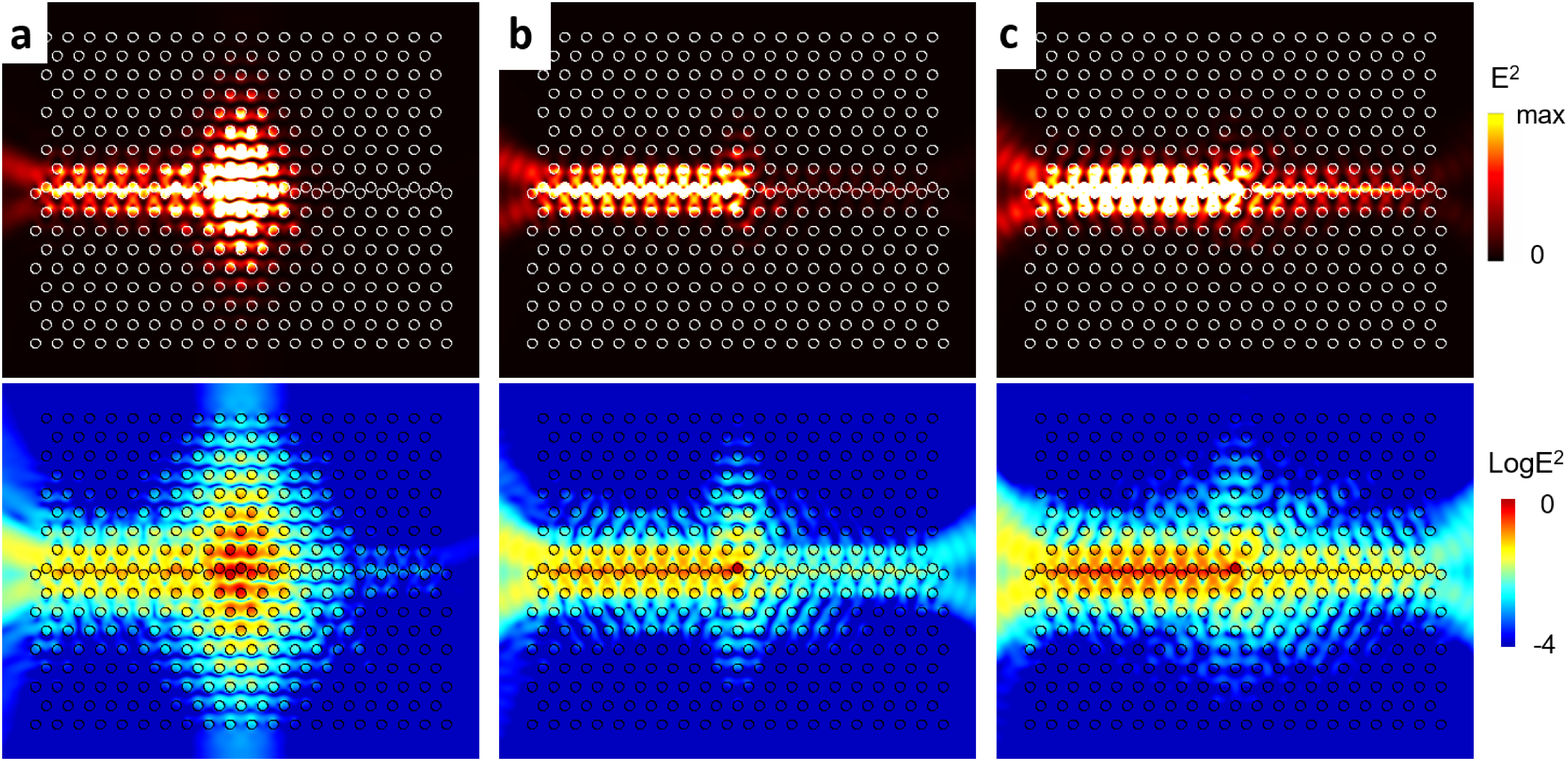}
  \caption{
Time-averaged electric field intensity distributions of the photonic topological edge modes propagating along the left direction in the straight interface with $r_b=0$. The upper is the electric intensity distributions with linear scale and the lower is the same with logarithmic scale excited by the CW chiral source with (a) 1500 nm, (b)1560 nm, and (c) 1610 nm, respectively.
}
  \label{fieldstr}
\end{figure}

In order to investigate the dispersion properties of the photonic topological edge mode qualitatively, we calculated the time-averaged electric field intensity distribution of the edge mode with different wavelength by 3-D FDTD simulation. Figure ~\ref{fieldstr} shows the xy-cut view of the intensity distribution when the CW chrial source is located at the center of the upper air hole in the interface. According to Fig. ~\ref{interface}(a) and Fig.~\ref{map}(b), directionality at $\lambda=1500$nm is almost perfect with a moderate group velocity. And, at $\lambda=1560$nm the field intensity is slightly propagating along the right direction so the directionality is still very high about 0.9. However, at $\lambda=1610$ nm the directionality decreases further, because of the band edge effect of the Tri-PhC mode. At the lower figures in Fig. ~\ref{fieldstr}(c), the field intensity along the left interface is stronger than that with different wavelength, which indicates that the group velocity of the edge mode is smaller than other modes. And also the electric field distribution near the CW chiral source looks larger than that with different wavelength. Especially, the evanescent field along K-direction is observed which means that the edge mode decays exponentially with the field profile of Tri-PhC mode at K-point. 

\begin{figure}[tbp!]
  \includegraphics[width=0.8\textwidth]{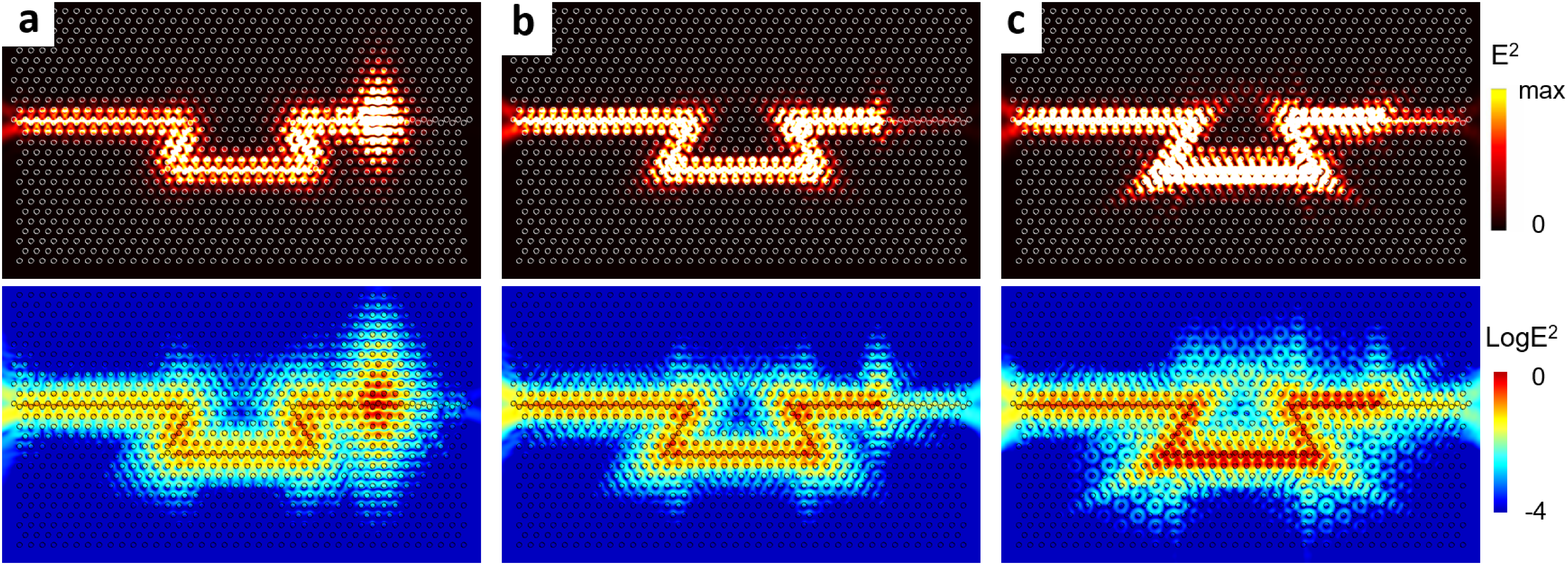}
  \caption{
Time-averaged electric field intensity distribution of the photonic topological edge modes in the {\rotatebox[origin=c]{180}{$\Omega$}}-shape interface with $r_b=0$. The upper is the electric intensity distribution with linear scale and the lower is the same with logarithmic scale at (a) 1500 nm, (b)1560 nm, and (c) 1610 nm.
}
  \label{fieldzig}
\end{figure}

We also calculated time-averaged electric field intensity distribution of the edge modes in the {\rotatebox[origin=c]{180}{$\Omega$}}-shape interfaces for checking the bending loss. Figure ~\ref{fieldzig} shows the xy-cut view of the intensity distribution when the CW chrial source is located at the right-middle of the upper air hole in the interface. At   $\lambda=1500$nm, the directionality is almost perfect which means no reflection observed as shown in Fig. ~\ref{fieldzig}(a). However, as the wavelength increases, the directionality decreases gradually and the reflection increases as shown in Fig. ~\ref{fieldzig}(b, c).   

\begin{figure}[tbp!]
  \includegraphics[width=0.8\textwidth]{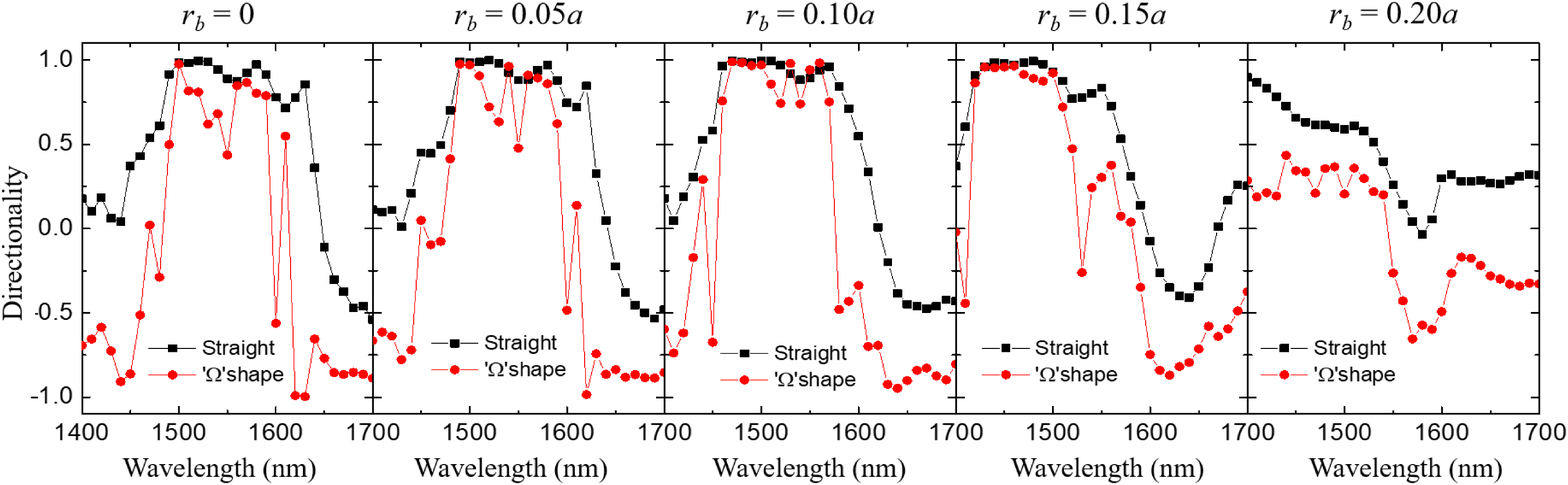}
  \caption{
 Directionality of the excited edge mode in the straight and {\rotatebox[origin=c]{180}{$\Omega$}}-shape interfaces with various $r_b$.  
}
  \label{interfaceS}
\end{figure}

In order to investigate the reflection at the bending of the interface quantitatively, the directionality of the edge modes in the {\rotatebox[origin=c]{180}{$\Omega$}}-shape interfaces with various  $r_b$ is compared with the directionality in the straight interfaces as shown in Fig. ~\ref{interfaceS}. At $r_b=0$, the directionality of edge mode in the {\rotatebox[origin=c]{180}{$\Omega$}}-shape interface is slightly smaller than that in the straight interface. In view of four 120$^\circ$ bending structures, the reflection at the single bending is very small. Especially, as $r_b$ increases, the directionality of the edge mode increases gradually, and almost the same in the straight interface, which means the no reflection at the bending of the interface. However, with further increasing $r_b/a$ over 0.15, the directionality becomes small. From this, we expect that there is an optimum $r_b/a$ value which has no reflection with high directionality.     

\end{document}